\documentclass[showpacs,preprintnumbers,amsmath,amssymb,twocolumn,superscriptaddress,prb]{revtex4}
\usepackage{graphicx}
\usepackage{amsfonts}
\usepackage{amsmath, amsthm, amssymb}
\usepackage{dsfont}
\usepackage{times}

\newcommand{\e}[1]{\mathrm{e}^{#1}}

\newcommand{\pdiff}[2]{\frac{\partial #1}{\partial #2}}

\newcommand{\ie}{\textit{i.e. }}
\newcommand{\eg}{\textit{e.g. }}
\newcommand{\etal}{\emph{et al.}}
\def\i{\mathrm{i}}

\begin{document}
\title[0-$\pi$ phase shifts in Josephson junctions as a signature for the $s_{\pm}$-wave pairing state]{0-$\pi$ phase shifts in Josephson junctions as a signature for the $s_{\pm}$-wave pairing state}
\author{Jacob Linder}
\affiliation{Department of Physics, Norwegian University of
Science and Technology, N-7491 Trondheim, Norway}
\author{Iver B. Sperstad}
\affiliation{Department of Physics, Norwegian University of
Science and Technology, N-7491 Trondheim, Norway}
\author{Asle Sudb{\o}}
\affiliation{Department of Physics, Norwegian University of
Science and Technology, N-7491 Trondheim, Norway}

\date{Received \today}
\begin{abstract}
\noindent We investigate Josephson junctions with superconducting ferropnictides, both in the diffusive and ballistic limit. 
We focus on the proposed $s_\pm$-wave state, and find that the relative phase shift intrinsic to the $s_\pm$-wave state 
may provide 0-$\pi$ oscillations 
in the Josephson current. This feature can be used to discriminate this pairing state from the conventional $s$-wave symmetry. 
The 0-$\pi$ oscillations appear both as a function of the ratio of the interface resistances for each band and, more 
importantly, as a function of temperature, which greatly aids in their detection. 
\end{abstract}

\maketitle
The discovery of high-$T_c$ superconductivity in the ferropnictides \cite{kamihara_jacs_08} has triggered an 
avalanche of investigations 
(see the reviews \cite{reviews} and references therein) from a broad range of communities 
in condensed matter physics. 
A crucial issue which remains unresolved is the nature of the superconducting 
order parameter (OP) symmetry in ferropnictide superconductors. This topic is 
particularly intriguing since the ferropnictides feature a multiband
Fermi-surface where the Cooper-pairs may reside. 
\par
In order to identify the symmetry of the superconducting order parameter (OP), several recent experimental 
studies \cite{pcs_nodal, pcs_gapped} utilized the method of point-contact spectroscopy in order to study the symmetry 
of the superconducting OP in the ferropnictides. The findings were, however, not easily reconcilable. Using an extended 
Blonder-Tinkham-Klapwijk (BTK) theory \cite{blonder_prb_82} to fit their data, some groups \cite{pcs_nodal} found a 
zero-bias conductance peak, indicative of a nodal order parameter such as $d$-wave. However, other groups \cite{pcs_gapped} 
interpreted their data in terms of one or more nodeless OPs, such as $s$-wave.
\par
One of the leading candidates for the pairing symmetry is the so-called $s_\pm$-wave state proposed in 
Refs. \cite{mazin_prl_08, seo_prl_08}. This pairing symmetry consists of two $s$-wave order parameters for 
the electron-like and hole-like Fermi surfaces that differ in sign. Some progress has been made in mapping 
out the ramifications of the $s_\pm$-wave symmetry to quantum transport properties of the ferropnictides 
\cite{linder_rapid_09, tsai_arxiv_08, boundstates}. For instance, it has been predicted 
that subgap surface states should appear in the presence of interband scattering \cite{boundstates}. Unfortunately, 
such subgap surface states are not unique for the $s_\pm$-wave state, and do not provide unambiguous evidence 
for this pairing symmetry.
\par
To shed more light on the pairing symmetry in the ferropnictide superconductors, 
we present results for both the proximity effect and the Josephson current in hybrid structures involving 
normal metal elements and superconducting ferropnictides. The motivation for this is that both of these phenomena are 
expected to produce valuable information about the pairing state in the superconductor. We take into account the 
intrinsic multiband nature of this material class and include results for the diffusive limit of transport, in contrast to previous theoretical works on these systems.
\par
For Josephson junctions with conventional superconductors ($s$-wave), it is well-known that the supercurrent decays in 
a monotonous fashion as a function of both temperature and interlayer width, when the material separating the superconductors 
is non-magnetic. If the interlayer is ferromagnetic, the current oscillates and goes to zero at certain critical 
widths and temperatures. This phenomenon is known as 0-$\pi$ oscillations \cite{bergeret_rmp_05}, and serves as a 
signature of either ferromagnetic correlations or nodal OPs, such as $d$-wave, present in the Josephson junction. 
\par
In this Rapid Communication, we show that the aforementioned prerequisites for 0-$\pi$ oscillations are rendered unnecessary in the presence 
of an $s_\pm$-wave pairing state. We find that 0-$\pi$ oscillations may occur in a Josephson junction consisting 
of a conventional $s$-wave superconductor and a $s_\pm$-wave superconductor separated by a normal (non-magnetic) 
interlayer, thus in the complete absence of any ferromagnetic elements or nodal superconducting OPs. This effect 
is explained in terms of the relative phase shift between the bands in the $s_\pm$-wave superconductor, and 
constitutes a signature of the $s_\pm$-wave state which can be probed in experiments. In fact,
using such an observation in conjunction with other experiments that report a nodeless OP, ruling out $d$-wave
pairing, would strongly support the realization of a $s_\pm$-wave state. Our results are qualitatively independent of the interband scattering strength, and are induced solely by the $s_\pm$-wave symmetry. This renders our prediction more robust than recent proposals regarding subgap bound states as probes for the $s_\pm$-wave state, which rely heavily on substantial interband scattering.
\par
We will employ the quasiclassical theory 
of superconductivity in form of the Usadel \cite{usadel} equation and the accompanying Kupriyanov-Lukichev boundary 
conditions \cite{kupluk} modified for a multiband situation \cite{brinkman_prb_04}. The quasiclassical approach is justified under the condition that the Fermi energy is much larger than the superconducting gap and the impurity scattering self-energy, which should be a safe assumption for the ferropnictides. The notation and conventions 
of Ref. \cite{linder_prb_08} will be used in what follows. For equilibrium situations, it suffices to consider 
the retarded part of the matrix Green's function, $\hat{g}$, which is parameterized conveniently by the quantity 
$\theta_\sigma^N$, $\sigma=\uparrow,\downarrow$. The Green's function satisfies $\hat{g}^2 = \hat{1}$ and consists of entries with 
$c_\sigma^N = \text{cosh}(\theta_\sigma^N)$ and $s_\sigma^N = \text{sinh}(\theta_\sigma^N)$ as measures of the proximity effect induced by the multiband superconductor. In this parameterization, the Usadel equation \cite{usadel} is obtained as $D_N \partial_x^2 \theta_\sigma^N + 2\i\varepsilon s_\sigma^N = 0,$
where $D_N$ is the diffusion coefficient in the normal metal and $\varepsilon$ is the quasiparticle energy. In the 
superconducting region, we use the bulk Green's functions $\hat{g}_\lambda$ \cite{bergeret_rmp_05, linder_prb_08} for each band as denoted by the index $\lambda = (1,2)$, with belonging 
gaps $\Delta_\lambda = |\Delta_\lambda|\e{\i\varphi_\lambda}$. The unique feature of the $s_\pm$-wave state is that 
the relative phase between the bands is $\pi$, i.e. $\varphi_1=\varphi$ and   
$\varphi_2 = \varphi + \pi$, where $\varphi$ is the superconducting phase associated with the broken U(1) gauge symmetry.
\par
The Usadel equation must be supplemented with boundary conditions at the interface of the superconducting region. Under 
the assumption of a low interface transparency, we may employ generalized Kupriyanov-Lukichev boundary conditions which 
for an N$\mid$$s_\pm$-wave interface at $x=d_N$ take the form $d_N\hat{g}_N\partial_x \hat{g}_N|_{x=d_N} =$ $\sum_\lambda \frac{1}{\gamma_\lambda} [\hat{g}_N, \hat{g}_\lambda]|_{x=d_N}$ where $d_N$ is the thickness of the normal metal layer while $\gamma_\lambda = R_B^\lambda/R_N$. Here, $R_N$ is the 
resistance of the normal metal region, while $R_B^\lambda$ is the effective barrier resistance for band $\lambda$. 
At $x=0$, we have $\partial_x \theta_\sigma^N=0$, corresponding to zero outgoing current at the insulating/vacuum 
interface. 
\par
\begin{figure}[t!]
\centering
\resizebox{0.45\textwidth}{!}{
\includegraphics{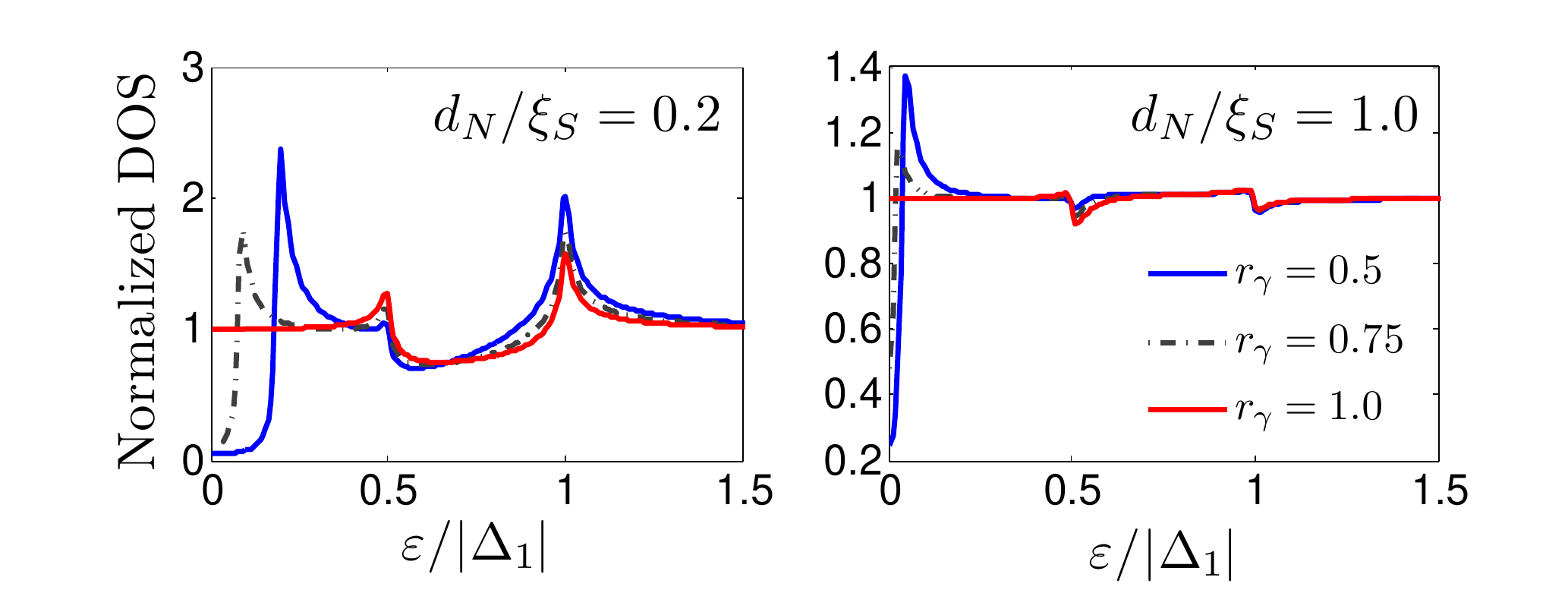}}
\caption{(Color online) Plot of the density of states (DOS) at $x=0$ (at the N$\mid$I interface) for a thin ($d_N/\xi_S=0.2$) and 
thick ($d_N/\xi_S=1.0$) normal metal region. We have set $r_\Delta=0.5$ and considered several values of $r_\gamma$. }
\label{fig:DOS_diffusive_prl}
\end{figure}

Let us first briefly investigate the full proximity effect regime in a N$\mid$$s_\pm$ junction by solving the Usadel equation numerically with its boundary conditions. The normalized density of states (DOS) reads $N(\varepsilon)/N_0 = \sum_\sigma \text{Re}\{c_\sigma^N\}$/2. There are three parameters that are free to vary in 
our theory. One is the thickness of the normal metal layer $d_N/\xi_S$, where $\xi_S = \sqrt{D_N/|\Delta_1|}$. The two others are the ratio between the gaps and the 
ratio between the barrier parameters, defined respectively as $r_\Delta = |\Delta_2/\Delta_1|$ and 
$r_\gamma=\gamma_2/\gamma_1$. In Fig. \ref{fig:DOS_diffusive_prl}, we contrast the thin junction case 
$d_N/\xi_S\ll1$ with a thick junction $d_N/\xi_S = 1$ for a representative choice of parameters. 
We fix $r_\Delta=0.5$ and plot the DOS in the N region at $x=0$ for several 
values of $r_\gamma$, with $\gamma_1=5$ corresponding to a low barrier transparency. There are in general three peaks in the energy-resolved DOS. Two of these 
peaks pertain to the bulk gaps of the $s_\pm$ superconductor, while the third demarcates the opening of 
a minigap in the spectrum. This is qualitatively the same as what would be expected for a multiband 
superconductor with a conventional $s$-wave symmetry, such as MgB$_2$ \cite{brinkman_prb_04}. 

Therefore, the proximity effect and its impact on the DOS does not appear to provide a 
unique diagnostic tool in order to distinguish $s_\pm$-wave symmetry from ordinary $s$-wave symmetry. We thus turn our attention to the Josephson coupling for $s_\pm$-wave superconductors as a possible mean to reveal this symmetry. 
To this end, we will consider a $s$-wave$\mid$N$\mid$$s_\pm$-wave junction, where the $s$-wave gap is given by 
$\Delta_s = |\Delta_s|\e{\i\varphi_s}$, and assume a weak proximity effect which allows us to linearize the Usadel equation and proceed analytically, facilitating the interpretation of the obtained results. Also, the linearized approach is expected to yield excellent results in the experimentally relevant low-transparency case. The supercurrent is given 
by $I_J \sim \int^\infty_{-\infty} \text{d}\varepsilon \text{Tr}\{ \hat{\rho}_3 (\hat{g}\partial_x\hat{g})^\text{K}\}$, 
where $\hat{\rho}_3=\text{diag}(1,1,-1,-1)$ and 'K' denotes the Keldysh component of the Green's function 
\cite{bergeret_rmp_05}. After solving the Usadel equation, one may insert $\hat{g}$ into the above equation for the supercurrent. We find the following expression for 
the normalized zero-temperature Josephson current: 
\begin{align}\label{eq:jos}
I_J = I_0\sin\Delta\varphi,\; I_0 = \int^\infty_0 \text{d}\varepsilon \text{Re}\{\mathcal{R}\mathcal{L}/[\i kd\sin(kd)]\},
\end{align}
where $\mathcal{L} 
= \sum_\lambda \delta_\lambda^L \mathcal{F}_\lambda^L/\gamma_\lambda$, $\mathcal{R} = \sum_\lambda \delta_\lambda^R
\mathcal{F}_\lambda^R/\gamma_\lambda$. 
Here, $\Delta\varphi = \varphi - \varphi_\text{s}$ is defined as the phase difference between band $\lambda = 1$ in the right superconductor 
and the left superconductor, $k=\sqrt{2\i\varepsilon/D_N}$, while $\mathcal{F}_\lambda^{L,R}$ describe the anomalous 
Green's functions on the left/right side of the junction. These are 
proportional to the off-diagonal entries in the bulk Green's functions for the superconductors, which 
have the form $\mathcal{F}_\lambda^{L,R} \propto  s_\lambda^{L,R}$. We defined $\delta_{\lambda=1} = 1$ and
$\delta_{\lambda=2}=-1$. Note that the above expressions are valid for both a $s$-wave 
and $s_\pm$-wave superconductor on either side of the diffusive normal metal, which is why we have included the band index also on the left side. In the $s$-wave case, 
we have $\mathcal{F}_\lambda = \delta_\lambda \text{sinh}[\text{arctanh}(|\Delta_s|/\varepsilon)]$, 
while in the $s_\pm$-wave case we have $\mathcal{F}_\lambda = \text{sinh}[\text{arctanh}(|\Delta_\lambda|/\varepsilon)]$.

\begin{figure}[t!]
\centering
\resizebox{0.45\textwidth}{!}{
\includegraphics{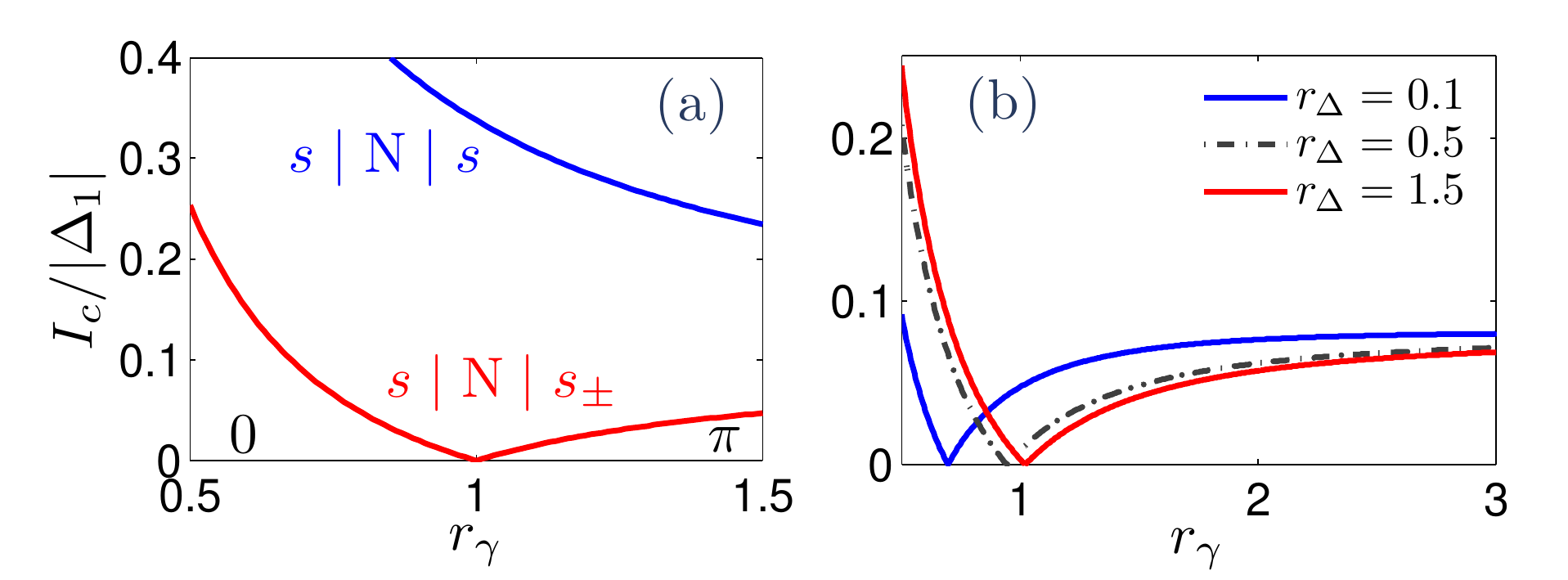}}
\caption{(Color online) (a) Plot of the critical current for an $s$-wave$\mid$N$\mid$$s$-wave and $s$-wave$\mid$N$\mid$$s_\pm$-wave 
junction, using $r_\Delta=1.0$ and $|\Delta_s/\Delta_1|=1.0$. (b) Plot of the critical current 
in the $s$-wave$\mid$N$\mid$$s_\pm$-wave case, using $|\Delta_s/\Delta_1|=0.5$. In both (a) and (b), 
we have set $d_N/\xi_S=1.0$.}
\label{fig:Josephson_diffusive}
\end{figure}

We now solve Eq. (\ref{eq:jos}) numerically to obtain the Josephson critical current, corresponding to $I_c = |I_0|$ which is the relevant 
quantity measured experimentally. In Fig. \ref{fig:Josephson_diffusive}(a), we plot the critical current as a function of the ratio 
between the interface barriers for each band, $r_\gamma$, for both a $s$-wave$\mid$N$\mid$$s$-wave and $s$-wave$\mid$N$\mid$$s_\pm$-wave 
junction. In the former case, the current decays monotonously as is well-known. However, the situation is very different when we replace, 
say, the right $s$-wave superconductor with an $s_\pm$-wave state. \textit{The current now displays 0-$\pi$ oscillations, even in 
the complete absence of any ferromagnetic elements}. This is very different from the conventional $s$-wave case, where a ferromagnetic
element is required in order to induce the 0-$\pi$ oscillations. Thus, experimental observation of such 0-$\pi$ oscillations
in a Josephson junction with ferropnictides would provide a strong indication of the presence of an $s_\pm$-wave state. In Fig.
\ref{fig:Josephson_diffusive}(b), we give results up to large $r_\gamma$ for the $s$-wave$\mid$N$\mid$$s_\pm$-wave case. As seen, 
the current saturates after the 0-$\pi$ oscillation since $r_\gamma\gg1$ means that one of the band interface transparencies 
tends to zero and does not contribute to transport.
\par
The appearance of the 0-$\pi$ oscillations in the current may be understood as follows. The transport of charge 
in an $s$-wave$\mid$N$\mid$$s_\pm$-wave junction takes place both through inter- and intraband channels, as may be inferred directly 
by observing that the product $\mathcal{L}\mathcal{R}$ in Eq. (\ref{eq:jos}) produces precisely such terms. Due to the relative 
phase shift of $\pi$ between the two bands in the $s_\pm$-wave state, these contributions to the critical current have opposite signs. For simplicity, consider the case where all gap magnitudes are equal in the Josephson junction, 
$|\Delta_\lambda|=|\Delta_s|$, which leads to equal anomalous Green's functions $\mathcal{F}$ on both sides of the junction. 
We then have $\mathcal{L}\mathcal{R} = \mathcal{F}^2(1/\gamma_1^2 - 1/\gamma_2^2)$ in Eq. (\ref{eq:jos}), which is clearly 
seen to change sign at $r_\gamma=1$. This does not occur in a conventional $s$-wave superconductor, where there is no 
relative phase shift. {\it The basic mechanism behind the 0-$\pi$ oscillations is thus that variations in the barrier 
parameters $\gamma_\lambda$ for the bands will lead to either a dominant contribution between bands with no phase 
shift relative each other or bands with order parameters that differ in sign.} 
\par
Let us also consider the ballistic limit, to show that the mechanism for the 0-$\pi$ oscillations persists in clean samples. The only other change in 
the physical system under consideration is that we replace the normal interlayer with a thin insulating barrier (I), which in the BTK 
approach introduces the dimensionless barrier strengths $Z_\lambda$. In this manner, we can parameterize the relative barrier resistance 
in an analogous manner as with $r_\gamma$ in the diffusive case by introducing $r_Z = Z_2/Z_1$. We construct and solve the full $4 \times 4$
Bogoliubov-de Gennes equation for the two-band system, where we for generality also include coupling between the two bands parameterized 
by the interband coupling strength $\alpha$. This yields in general four current-carrying Andreev Bound States (ABS) 
$E_\lambda^\pm(\Delta\varphi)$. The Josephson current for this $s$-wave$|$I$|$$s_\pm$-wave Josephson junction is then found in the 
ordinary way from \cite{kwon} $I_J = 2 e \sum_{i=1}^4 \pdiff{E_i}{\varphi} f(E_i),$where $E_i$ denotes the four ABS and $f(E)$ 
is the Fermi-Dirac distribution function.

\begin{figure}[t!]
	\centering
	\resizebox{0.45\textwidth}{!}{
	\includegraphics{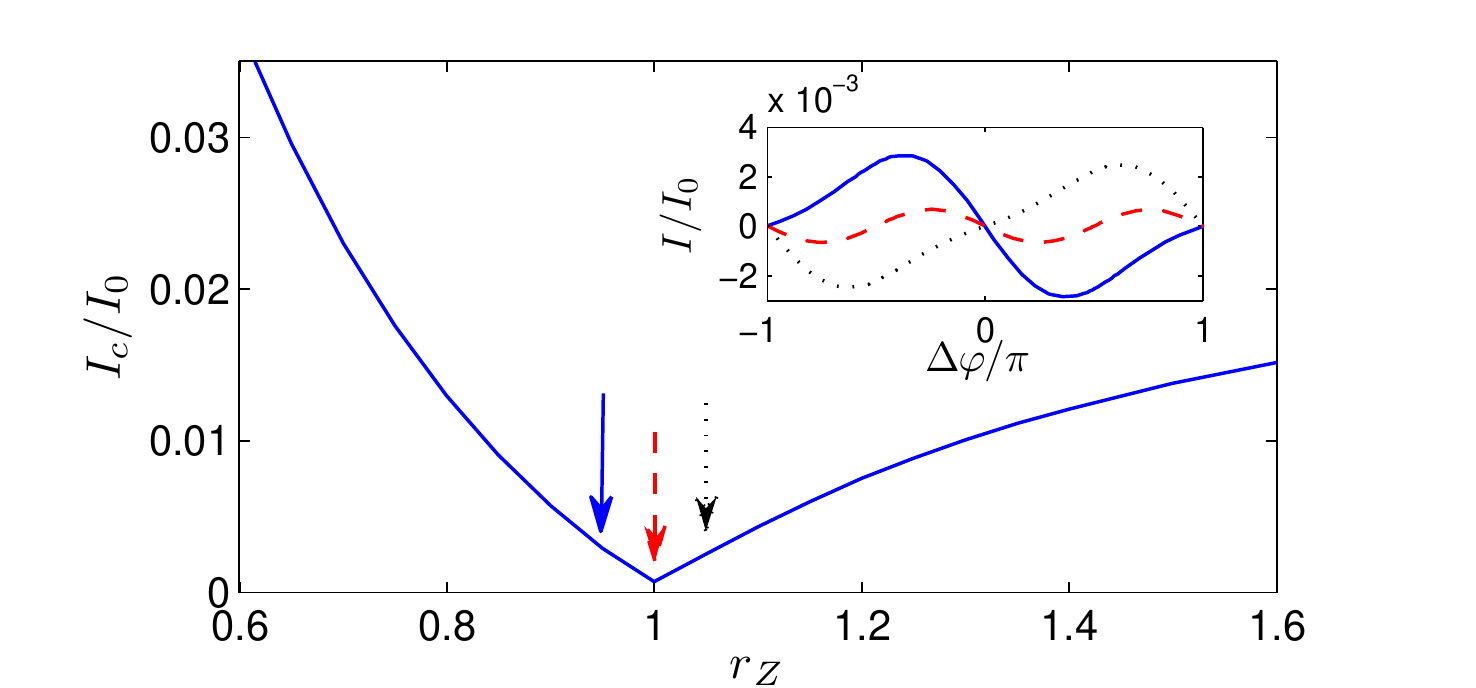}
	} \caption{(Color online) Critical current for a ballistic $s$-wave$|$I$|$$s_\pm$-wave Josephson junction as a function of the relative barrier strength $r_Z$. Interband coupling is neglected, and we have set $Z_1 = 6$, $T = 0$, and $|\Delta_\lambda| = |\Delta_s|$. \textit{Inset:} current-phase-relation for selected values of $r_Z$, as indicated by the arrows in the main figure. }
	\label{fig:Ic_vs_rZ}
\end{figure}

To present an explicit illustration of the mechanism of 0-$\pi$-oscillations in a $s_\pm$-system in the ballistic limit, we proceed analytically for the special case of $\alpha = 0$. Here, we have for simplicity assumed that $|\Delta_\lambda| = |\Delta_s| \equiv |\Delta|$. This gives solutions for the ABS on the well-known\cite{kwon} form $E_1^\pm = \pm |\Delta| \sqrt{1 - D_1 \sin^2(\Delta\varphi/2)}$ and $E_2^\pm = \pm |\Delta| \sqrt{1 - D_2 \cos^2(\Delta\varphi/2)}$, with $D_\lambda = 4 / (4 + Z_\lambda^2)$. At $T = 0$, the above expression for the Josephson current yields in the tunneling limit $I_J = I_1 \sin{\varphi}$, with $I_1 =(D_1 - D_2) I_0 / 4$ and $I_0 = 2 e |\Delta|$. It is obvious that for $Z_2 < Z_1$ one will have $I_1 < 0$, \ie the system being in the $\pi$ state, as explained for the diffusive case. As shown in Fig. \ref{fig:Ic_vs_rZ}, the crossover point above which the $\lambda = 1$ contribution dominates instead is $r_Z = 1$. Notice however that the current does not vanish entirely at the crossover point due to a second harmonic component in the current-phase-relation (as shown in the inset of Fig. \ref{fig:Ic_vs_rZ}) which dominates close to the transition point. This is demonstrated explicitly by taking the approximation to the next order in the limit $Z_2 = Z_1$, which yields $I_J = I_2 \sin(2\Delta\varphi)$, with $I_2 = - I_0 D_\lambda^2/ 16$. We note that this non-sinusoidality of the current-phase-relation was absent in the diffusive case since the linearized Usadel equation corresponds to a first order approximation in the interface resistance. We also emphasize
that in this treatment, interband coupling is not essential for the
occurrence of the 0-$\pi$-transition. However, we have verified
numerically that the results of Fig. \ref{fig:Ic_vs_rZ} are
qualitatively valid also for $\alpha > 0$, so that the predicted
experimental signature should be equally distinct for strong interband
coupling.
\par
From the analysis above, 
it is seen that the crucial ingredient for the observation of the 0-$\pi$ oscillations is having different barrier 
parameters for each band $\lambda$, or alternatively different probabilities for Cooper-pair tunnelling. As suggested 
in Ref. \cite{tsai_arxiv_08}, these probabilities may be artificially altered by selecting materials with appropriate 
Fermi surfaces. Different Fermi vector-mismatches would then lead to different tunneling probabilities. In our case, 
the size of the Fermi surface of the diffusive normal metal region could be modified by doping. Thus, whereas 0-$\pi$ 
oscillations in S$\mid$F$\mid$S junctions can be seen as a function of the width $d_F$ of the ferromagnetic layer 
\cite{kontos_prl_02}, necessitating the fabrication of several samples with different widths, the present scenario 
requires fabrication of several samples with the doping level in the normal metal varying in a systematic way. We 
note that it was also observed in Ref. \cite{tsai_arxiv_08}, although in the context of a superconducting 
$s$-wave$\mid$$s_\pm$-wave$\mid$$s$-wave trilayer, that a $\pi$-junction could be fabricated in a similar manner.

\begin{figure}[t!]
\centering
\resizebox{0.45\textwidth}{!}{
\includegraphics{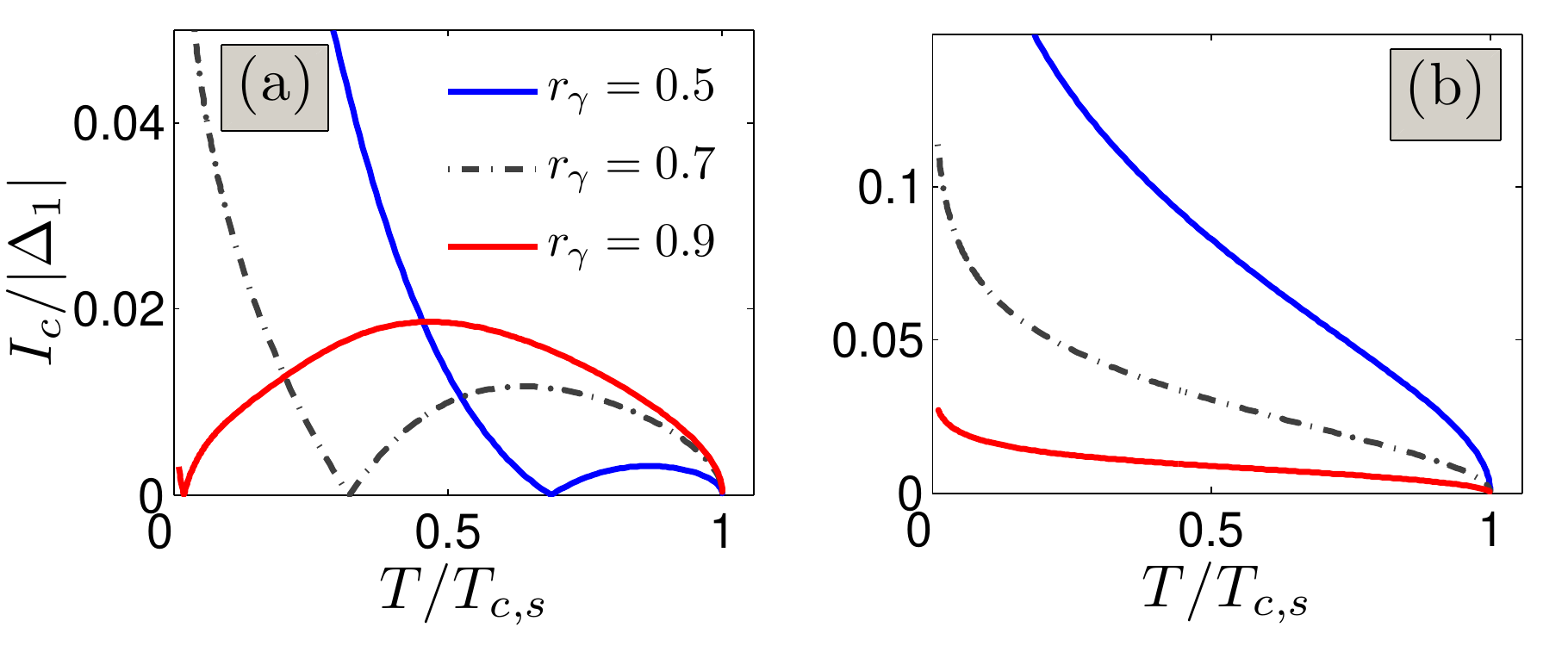}}
\caption{(Color online) Plot of the critical current as a function of temperature for an $s$-wave$\mid$N$\mid$$s_\pm$-wave junction, using $d_N/\xi_S=1.0$ and $|\Delta_s/\Delta_1|=0.5$. In (a), we have $r_\Delta=0.3$ while in (b) $r_\Delta=1.3$. }
\label{fig:Josephson_temp}
\end{figure}

Although the above procedure is in principle feasible, it is very challenging to quantitatively relate the Fermi-vector 
mismatch directly to the parameter $r_\gamma$. However, we find that \textit{the 0-$\pi$ oscillations also occur as a 
function of temperature in the diffusive limit}, thus constituting an alternative, and simpler, approach to the recipe sketched above for 
altering $r_\gamma$. Assuming a Bardeen-Cooper-Schrieffer (BCS) temperature-dependence 
for the gaps, with a critical temperature $T_{c,\lambda}=T_c$ for the $s_\pm$-wave superconductor and $T_{c,s}$ for 
the $s$-wave superconductor \footnote{Note that at $T\neq0$, an additional term $\text{tanh}(\beta\varepsilon/2)$ appears 
in the integrand of Eq. (\ref{eq:jos}), where $\beta=1/T$.}, we plot the results in Fig. \ref{fig:Josephson_temp}. As 
seen, 0-$\pi$ oscillations appear 
as a function 
of temperature for a wide range of interface parameters $r_\gamma$. For large values of $r_\Delta$, a normal monotonous 
decay of the critical current is seen. Although the exact relation between $r_\gamma$ and $r_\Delta$ which renders possible 
the 0-$\pi$ oscillations is difficult to extract analytically from Eq. (\ref{eq:jos}), the basic mechanism is 
nevertheless the same as the one explained previously. From Fig. \ref{fig:Josephson_temp}, we see that the absence of 0-$\pi$ oscillations not necessarily rules out that $s_\pm$ state, whereas the presence of them rules out the $s$-wave state.
\par
Finally, we point out that very strong impurity interband scattering $\Gamma$ would eventually suppress the critical temperature for the $s_\pm$ ground state \cite{bang_prb_09}. The difference between the DOS on the hole and electron Fermi pockets would determine how fast the suppression rate increases with $\Gamma$ as compared to \eg a $d$-wave scenario. For intraband scattering, however, the $s_\pm$ state is protected by Anderson's theorem. In our model, we have incorporated interband scattering only near the interface. A further extension of the model considered here could be to incorporate magnetic correlations in the $s_\pm$ state and also investigate strong interband scattering in the bulk of the superconductor to see how it affects the transport properties \cite{stanev_prb_08}, although we expect that they would remain qualitatively the same as reported here since the basic mechanism for the 0-$\pi$ oscillations would remain intact.
\par
In summary, we have investigated the Josephson coupling properties of junctions with $s_\pm$-wave superconductors. In 
contrast to previous literature, we have here included results for both the ballistic and diffusive regimes. The relative phase shift of the bands intrinsic for the $s_\pm$-wave state 
leads to 0-$\pi$ oscillations in an $s$-wave$\mid$N$\mid$$s_\pm$-wave Josephson junction, even in the absence of any ferromagnetic 
elements. The mechanism behind these oscillations is a competition between the sign-dependent contribution of transport from 
different bands in the $s_\pm$-wave superconductor to the $s$-wave superconductor. The 0-$\pi$ oscillations are seen as a function 
of temperature, thus vastly facilitating the experimental testing of our predictions compared to methods that involve
changing the parameters of the model system. Our results may aid in identifying the possible existence of an $s_\pm$-wave pairing 
state in the superconducting ferropnictides. 
\par
\textit{Acknowledgments}. T. Yokoyama and Y. Tanaka are thanked for very useful discussions. J.L. and A.S. were 
supported by the Research Council of Norway, Grants No. 158518/431 and No. 158547/431 (NANOMAT), and Grant 
No. 167498/V30 (STORFORSK).

\end{document}